\documentclass[fleqn,usenatbib]{mnras}
\usepackage{float}
\usepackage{newtxtext}
\usepackage[slantedGreek]{newtxmath}
\usepackage[T1]{fontenc}
\DeclareRobustCommand{\VAN}[3]{#2}
\let\VANthebibliography\thebibliography
\def\thebibliography{\DeclareRobustCommand{\VAN}[3]{##3}\VANthebibliography}
\usepackage{graphicx}
\usepackage{amsmath}
\usepackage{xcolor,ulem}

\title[An off-axis jet seen in the TDE AT 2018hyz]{An off-axis relativistic jet seen in the long lasting delayed radio flare of the TDE AT 2018hyz}

\author[I. Sfaradi et al.]{
Itai Sfaradi,$^{1}$\thanks{E-mail: itai.sfaradi@mail.huji.ac.il}
Paz Beniamini,$^{2,3,4}$
Assaf Horesh,$^{1}$
Tsvi Piran,$^{1}$
Joe Bright,$^{5}$
Lauren Rhodes,$^{5}$
\newauthor
David R. A. Willians,$^{6}$
Rob Fender,$^{5}$
James K. Leung,$^{7,8,9}$
Tara Murphy,$^{7,9}$ and,
Dave A. Green$^{10}$
\\
\\
$^{1}$Racah Institute of Physics. The Hebrew University of Jerusalem. Jerusalem 91904, Israel \\
$^{2}$ Department of Natural Sciences, The Open University of Israel, P.O Box 808, Ra'anana 4353701, Israel \\
$^{3}$ Astrophysics Research Center of the Open University (ARCO), The Open University of Israel, P.O Box 808, Ra'anana 4353701, Israel \\
$^{4}$Department of Physics, The George Washington University, 725 21st Street NW, Washington, DC 20052, USA\\
$^{5}$ Astrophysics, Department of Physics, University of Oxford, Keble Road, Oxford OX1 3RH, UK \\
$^{6}$ Jodrell Bank Centre for Astrophysics, School of Physics and Astronomy, University of Manchester, Manchester M13 9PL, UK \\
$^{7}$ Sydney Institute for Astronomy, School of Physics, The University of Sydney, NSW 2006, Australia \\
$^{8}$ CSIRO Space and Astronomy, PO Box 76, Epping, NSW 1710, Australia \\
$^{9}$ ARC Centre of Excellence for Gravitational Wave Discovery (OzGrav), Hawthorn, Victoria, Australia
\\
$^{10}$ Astrophysics Group, Cavendish Laboratory, 19 J. J. Thomson Ave., Cambridge CB3 0HE, UK \\
}

\date{Accepted XXX. Received YYY; in original form ZZZ}

\pubyear{2023}

\begin{document}
\label{firstpage}
\pagerange{\pageref{firstpage}--\pageref{lastpage}}
\maketitle

\begin{abstract}
The Tidal Disruption Event (TDE) AT 2018hyz exhibited a delayed radio flare almost three years after the stellar disruption. Here we report new radio observations of the TDE AT 2018hyz with the AMI-LA and ATCA spanning from a month to more than four years after the optical discovery and 200 days since the last reported radio observation. We detected no radio detection from $30-220$ days after the optical discovery in our observations at $15.5$ GHz down to a $3\sigma$ level of $< 0.14$ mJy. The fast-rising, delayed, radio flare is observed in our radio data set and continues to rise almost $\sim 1580$ days after the optical discovery. We find that the delayed radio emission, first detected $972$ days after optical discovery, evolves as $t^{4.2 \pm 0.9}$, at $15.5$ GHz. Here, we present an off-axis jet model that can explain the full set of radio observations. In the context of this model, we require a powerful narrow jet with an isotropic equivalent kinetic energy $E_{\rm k,iso} \sim 10^{55}$ erg, an opening angle of $ \rm \sim 7^{\circ}$, and a relatively large viewing angle of $ \rm \sim 42^{\circ}$, launched at the time of the stellar disruption. Within our framework, we find that the minimal collimated energy possible for an off-axis jet from AT 2018hyz is $E_k \geq 3 \times 10^{52}$ erg. Finally, we provide predictions based on our model for the light curve turnover time, and for the proper motion of the radio emitting source.
\end{abstract}

\begin{keywords}
radio continuum: transients -- transients: tidal disruption events
\end{keywords}

\section{Introduction}
\label{sec: introduction}

A tidal disruption event (TDE) occurs when a star passes too close to a supermassive black hole (SMBH) and gets torn apart by the tidal forces exerted on it by the SMBH \citep{hills_1975,rees_1988}. About half of the disrupted star falls back to the SMBH and generates a multi-wavelength flare. While thermal emission is associated with the debris of the disrupted star (e.g., \citealt{rees_1988,cannizzo_1990,Komossa_2015,metzger_2016}), the origin of this emission, whether its the accretion onto the SMBH \citep{rees_1988,evans_1989,phinney_1989}, or some other process (e.g., internal shocks due to collisions in the debris stream \citealt{piran_2015,shiokawa_2015,liptai_2019,bonnerot_2020}) is still unclear.

Radio observations can play a key role in revealing the structure and evolution of any outflow ejected as a result of the TDE (see e.g \citealt{alexander_2020, van_velzen_2021}). Observations of thermal TDEs showed a wide variety of phenomena. For example, neutrino emission was associated with the three TDEs that were detected at radio wavelengths (AT 2019dsg; \citealt{stein_2021}, AT 2019fdr \citealt{reusch_2022}, and AT 2019aalc \citealt{van_velzen_2021_2019allc}), although the neutrino association of AT 2019dsg is still debated \citep{cendes_2021}. The prompt radio emission of ASASSN-14li was suggested to arise from either a sub-relativistic accretion-driven wind interacting with the circumnuclear material (CNM) \citep{alexander_2016}, or the unbound stellar debris traveling away from the SMBH \citep{krolik_2016}, or even a newly launched narrow jet \citep{van_velzen_2016}. A different subclass of TDEs are the ones showing relativistic outflows (e.g. SwiftJ1644+57; \citealt{Zauderer_2011,berger_2012,Zauderer_2013,cendes_2021_J1644}, SwiftJ2058+05; \citealt{cenko_2013,Pasham_2015,Brown_2017}, SwiftJ1112-82; \citealt{Brown_2017}, AT2022cmc; \citealt{Pasham_2023,rhodes_2023}). Radio observations of such a TDE, Swift J1644+57 \citep{Zauderer_2011,berger_2012}, uncovered a mildly relativistic collimated jet that then expanded and slowly decelerated, however, the origin of its radio spectra temporal evolution is still debated (see \citealt{berger_2012, piran_2013, Kumar2013, eftekhari_2018, Generozov2017, Beniamini_2023_J1644} for different models  explaining the observed increase in total energy by an order of magnitude).

A recent discovery of a new phenomenon that has been observed in several TDEs by now is the onset of late-time delayed radio flares (e.g ASASSN-15oi; \citealt{horesh_2021a}, iPTF16fnl; \citealt{horesh_2021b}, IGR J12580; \citealt{perlman_2022}, AT 2019azh; \citealt{sfaradi_2022}, and AT 2018hyz; \citealt{horesh_2018_18hyz,horesh_2022_18hyz,cendes_2022}). A simple explanation for a delayed radio flare is if an outflow was launched at a late stage (e.g. \citealt{horesh_2021a, cendes_2022}). Alternatively, relativistic collimated outflows, such as jets, can naturally explain delayed radio flares if the jet does not point toward the observer. The radio emission from them is highly sensitive to the viewing angle of the observer, and will increase once the jet  decelerates, and the emitting material enters the line of sight of the observer. This has been shown, for example, in studies of  gamma-ray bursts (GRBs; see e.g. \citealt{granot_2002}; \citealt{rossi_2002}; \citealt{Totani_2002}; \citealt{granot_2018}; \citealt{Beniamini2020}). The question of whether the underlying mechanism of delayed radio flares in TDEs is an off-axis jet that can only be observed in late times, or something else (e.g. delayed ejection of material due to a transition in accretion state) remains unanswered. Furthermore, it is also possible that both mechanisms (i.e. an off-axis jet and/or delayed outflow ejection) take place and manifest differently in different TDEs with delayed radio emission.

AT 2018hyz is a TDE at a redshift of $z=0.0457$ that was first discovered in optical by the All-Sky Automated Survey for Supernovae (ASAS-SN) on October 14, 2018, \citep{van_velzen_2021,gomez_2020}. \cite{cendes_2022} discovered a delayed radio flare, $972$ days after optical discovery, and claimed that the steep rise in radio flux density of this flare rules out an off-axis jet. However, recently \cite{matsumoto_2023} generalized the equipartition method for synchrotron self-absorbed radio sources \citep{pacholczyk_1970, scott_readhead_1977, chevalier_1998, barniol_duran_2013} to relativistic off-axis observed sources and have shown that an off-axis jet that was launched at the time of stellar disruption can explain the delayed emission from AT 2018hyz. In this work, we present new radio observations of the TDE AT 2018hyz and analyze all radio observations published so far in the context of this model. In \S\ref{sec: observations} we present our radio observation of this TDE and in \S\ref{sec: analysis} we model the radio emission from AT\,2018hyz as arising from an off-axis jet. \S\ref{sec: summary} is for summary and conclusions.

\section{Radio observations}
\label{sec: observations}

The first radio observation of AT 2018hyz with the Arcminute Microkelvin Imager - Large Array (AMI-LA; \citealt{zwart_2008}; \citealt{hickish_2018}) was conducted about $32$ days after optical discovery, with a central frequency of $15.5$ GHz, and resulted in flux density $3\sigma$ upper limit of $0.085$ mJy \citep{horesh_2018_18hyz}. \cite{horesh_2022_18hyz} reported the discovery of late-time radio emission from AT\,2018hyz with the Variable And Slow Transients survey (VAST; \citealt{Murphy_2021}) $1013$ days after optical discovery. The broadband observing campaign of this TDE with the Karl G. Jansky Very Large Array (VLA) revealed a delayed, late-time radio flare \citep{cendes_2022}. We report here radio observations of AT 2018hyz with AMI-LA prior to, and during, this delayed radio flare, between $32$ to $1578$ days after optical discovery. We also obtained a radio spectrum $1290$ days after optical discovery with the Australia Telescope Compact Array (ATCA; \citealt{wilson_2021_atca}). All upper limits quoted in this paper are at $3\sigma$ level.

\begin{table}
\centering
\begin{tabular}{cccc}
    \hline
    $\rm \Delta t \, \left[ days \right]$ & Frequency [GHz] & $F_{\rm \nu} \, \left[ mJy \right]$ & Telescope \\
    \hline
    $32$ & $15.5$ & $<0.085$ & AMI-LA \\
    $36$ & $15.5$ & $<0.072$ & AMI-LA \\
    $39$ & $15.5$ & $<0.069$ & AMI-LA \\
    $100$ & $15.5$ & $<0.13$ & AMI-LA \\
    $131$ & $15.5$ & $<0.093$ & AMI-LA \\
    $158$ & $15.5$ & $<0.11$ & AMI-LA \\
    $218$ & $15.5$ & $<0.14$ & AMI-LA \\
    $1242$ & $15.5$ & $3.83 \pm 0.77$ & AMI-LA \\
    $1245$ & $15.5$ & $4.06 \pm 0.81$ & AMI-LA \\
    $1247$ & $15.5$ & $3.93 \pm 0.79$ & AMI-LA \\
    $1254$ & $15.5$ & $4.11 \pm 0.82$ & AMI-LA \\
    $1262$ & $15.5$ & $4.07 \pm 0.81$ & AMI-LA \\
    $1274$ & $15.5$ & $4.46 \pm 0.89$ & AMI-LA \\
    $1290$ & $4.73$ & $7.98 \pm 0.80$ & ATCA \\
    $1290$ & $5.24$ & $7.66 \pm 0.77$ & ATCA \\
    $1290$ & $5.5$ & $7.68 \pm 0.77$ & ATCA \\
    $1290$ & $5.76$ & $7.59 \pm 0.76$ & ATCA \\
    $1290$ & $6.27$ & $7.35 \pm 0.74$ & ATCA \\
    $1290$ & $8.23$ & $6.58 \pm 0.66$ & ATCA \\
    $1290$ & $8.74$ & $6.37 \pm 0.64$ & ATCA \\
    $1290$ & $9.0$ & $6.31 \pm 0.63$ & ATCA \\
    $1290$ & $9.26$ & $6.28 \pm 0.63$ & ATCA \\
    $1290$ & $9.77$ & $6.05 \pm 0.60$ & ATCA \\
    $1290$ & $15.9$ & $4.81 \pm 0.48$ & ATCA \\
    $1290$ & $16.4$ & $4.64 \pm 0.46$ & ATCA \\
    $1290$ & $16.7$ & $4.58 \pm 0.46$ & ATCA \\
    $1290$ & $16.9$ & $4.49 \pm 0.45$ & ATCA \\
    $1290$ & $17.5$ & $4.46 \pm 0.45$ & ATCA \\
    $1291$ & $15.5$ & $5.0 \pm 1.0$ & AMI-LA \\
    $1326$ & $15.5$ & $5.35 \pm 1.07$ & AMI-LA \\
    $1392$ & $15.5$ & $6.65 \pm 1.33$ & AMI-LA \\
    $1466$ & $15.5$ & $7.72 \pm 1.54$ & AMI-LA \\
    $1496$ & $15.5$ & $8.69 \pm 1.74$ & AMI-LA \\
    $1578$ & $15.5$ & $10.2 \pm 2.0$ & AMI-LA \\
    \hline
\end{tabular}
\caption{AMI-LA and ATCA observations of AT 2018hyz. $\Delta t$ is the time in days since optical discovery, $F_{\rm \nu}$ is the flux density in mJy. Upper limits are $3\sigma$ image rms, and the flux density uncertainties are $20\%$ of the flux density for AMI-LA observations and $10\%$ of the flux density for ATCA observations. \label{tab: radio_table}}
\end{table}

\subsection{The Arcminute Microkelvin Imager - Large Array}
\label{subsec: AMI_observations}

AMI-LA is a radio interferometer comprised of eight, 12.8-m diameter, antennas producing 28 baselines that extend from 18-m up to 110-m in length and operate with a 5 GHz bandwidth, divided into eight channels, around a central frequency of 15.5 GHz. We reduced, flagged, and calibrated our observations using $\tt{reduce \_ dc}$, a customized AMI-LA data reduction software package (\citealt{perrott_2013}). Phase calibration was conducted using short interleaved observations of J1008+0621, while daily observations of 3C286 were used for absolute flux calibration. Additional flagging was performed using the Common Astronomy Software Applications (CASA; \citealt{2007ASPC..376..127M}). Images of the field of AT 2018hyz were produced using CASA task CLEAN in an interactive mode. We fitted the source in the phase centre of the images with the CASA task IMFIT and calculated the image rms with the CASA task IMSTAT. We estimate the error of the peak flux density to be a quadratic sum of the image rms, the error produced by CASA task IMFIT, and $20$ per cent calibration error (this is due to large variations in the phase calibrator flux density during the observations of this TDE). The flux density at each time is reported in Table \ref{tab: radio_table}.

\subsection{The Australia Telescope Compact Array}
\label{subsec: ATCA_observations}

We observed AT 2018hyz with the Australia Compact Telescope Array (project code C3363, PI: T. Murphy) from 08:00 to 14:00 UTC on 2022 April 26. Our observations were centred on frequencies 5.5, 9.0, 16.7, and 21.2 GHz, each with a bandwidth of 2048 MHz, with the array in the 1.5A configuration, which has a maximum baseline of 4.5 km.

We reduced the visibility data using standard routines in {\sc Miriad} \citep{Sault1995}. Prior to calibration, we had manually flagged radio-frequency interference by identifying outliers in the visibility data as both a function of time and frequency. In all frequency bands, PKS B1934$-$638 was used to calibrate both the bandpass response and flux-density scale, while PKS B1004$-$018 was used to calibrate the time-variable complex gains. Due to calibration issues, a reliable bandpass and flux-density scale could not be obtained at 21.2 GHz; therefore we have omitted the measurements from this band.

For all other frequency bands, we split the calibrated data into four sub-bands, each with a bandwidth of 512 MHz, to obtain higher spectral resolution; we then imaged each sub-band (512 MHz) and full band (2048 MHz). The imaging process, also implemented in {\sc Miriad}, involved weighting the visibilities with a robustness parameter of zero \citep{Briggs1995} prior to taking the inverse Fourier transform to obtain the dirty map of the sky brightness distribution. We obtained the final images by deconvolving the dirty map using the multi-frequency synthesis CLEAN algorithm \citep{Hogbom1974, Clark1980, Sault1994} and find a clear point-source detection of the source in each of our images. The reported flux densities were extracted from the images by fitting a point-source Gaussian model in the image plane using the {\sc Miriad} task {\sc imfit}. The flux density at each time is reported in Table \ref{tab: radio_table}.

\section{Radio data modelling}
\label{sec: analysis}

\begin{figure*}
\includegraphics[width=0.99\linewidth]{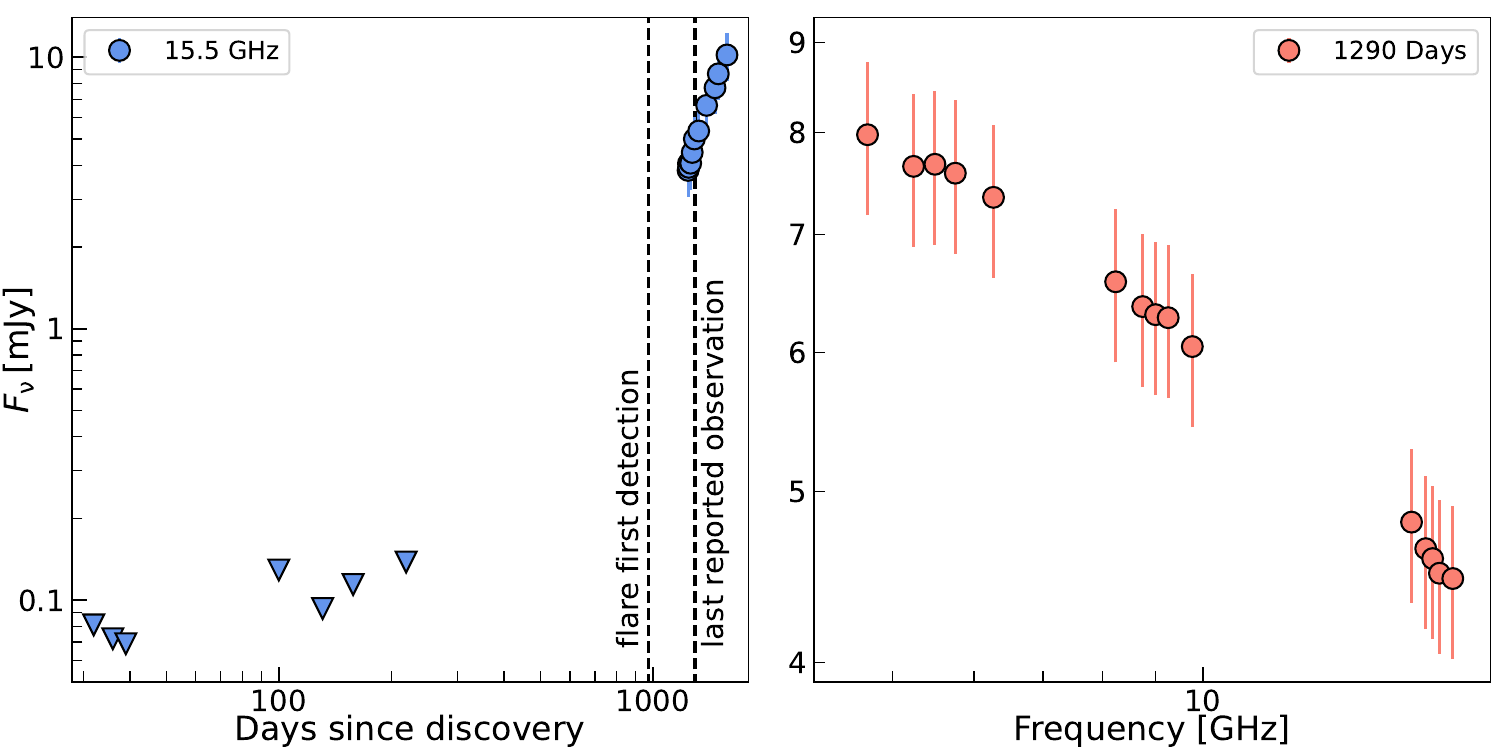}
    \caption{Left: $15.5$ GHz light curve of AT\,2018hyz with AMI-LA  (triangles mark $3\sigma$ upper limits). Also marked, for reference, on this plot are the times of the first radio detection of the flare, and of the last reported observation. Right: the radio spectrum, between $4.73$ to $17.5$ GHz, obtained with ATCA $1290$ days after optical discovery.}
    \label{fig: AMI_ATCA}
\end{figure*} 

Previously reported broadband radio observations of AT 2018hyz \cite{cendes_2022} showed no emission up to $\sim 970$ days after optical discovery, and then, a late time flare in all radio bands up to $\sim 1300$ days after optical discovery. The temporal evolution of the radio spectra showed an increase over time of the radio spectral peak flux density from $2.4$ to $8.8$ mJy, and a relatively constant peak frequency ($1.5$ to $3$ GHz) and electron power-law index ($2.1-2.3$). Equipartition analysis performed by \cite{cendes_2022} suggests a mildly relativistic outflow with $\beta = 0.25$ and $0.6$ ($\beta = \sqrt{1 - 1/\Gamma^2}$ is the source velocity normalised by the speed of light $c$) for spherical or jetted geometries, respectively, and a minimum kinetic energy of $\sim 6 \times 10^{49} \rm \, erg$.

\begin{figure}
\includegraphics[width=1\columnwidth]{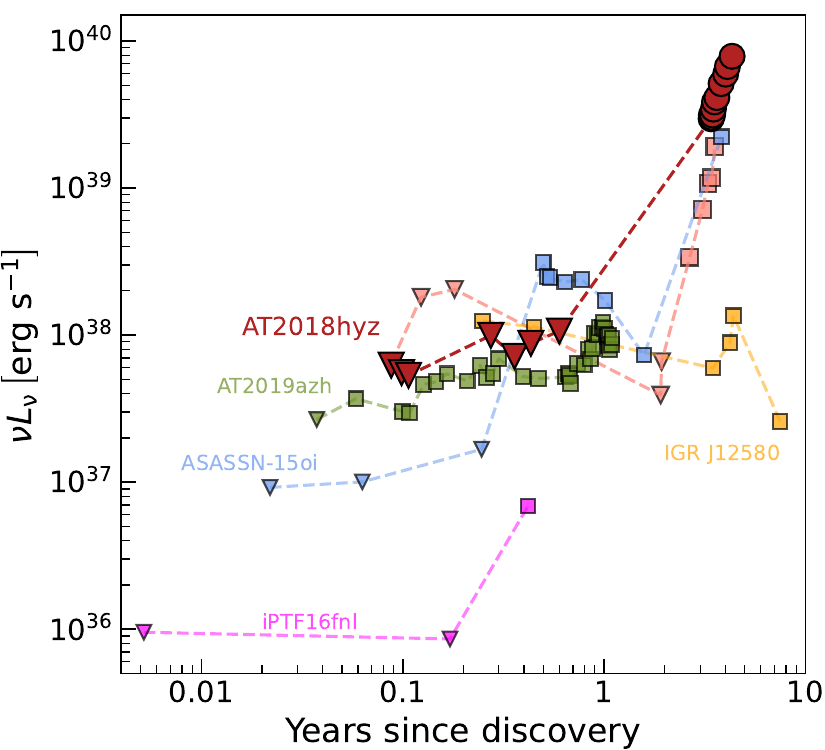}
    \caption{The radio luminosity as a function of time for known TDEs with delayed late-time radio flares (adapted from \citealt{alexander_2020}). The squares are of previously reported TDEs (ASASSN-15oi; \citealt{horesh_2021a}, iPTF16fnl; \citealt{horesh_2021b}, IGR J12580; \citealt{perlman_2022}, AT 2019azh; \citealt{sfaradi_2022}, and AT 2018hyz; \citealt{cendes_2022}). The circles are the $15.5$ GHz observations of AT 2018hyz. Triangles mark $3\sigma$ upper limits.}
    \label{fig: tde_comparison}
\end{figure}

Our $15.5$ GHz observations of the TDE AT 2018hyz (see left panel of Fig.~\ref{fig: AMI_ATCA}; $F_{\rm \nu}$ is the flux density in mJy) show a continuous rise in the radio emission $\sim 300$ days after the last observation reported by \cite{cendes_2022}. The $3\sigma$ upper limits first reported here between $100$ and $219$ days can be used to rule out an earlier flare down to $\sim 0.1$ mJy (see Table \ref{tab: radio_table}). This translates to a limit of $\nu L_{\nu} \leq 10^{38} \, \rm erg \, s^{-1}$, similar to what was seen, for example, in the TDE ASASSN-15oi (see Fig.~\ref{fig: tde_comparison} for a comparison with other known TDEs that have delayed radio flares). We also present here the radio spectrum obtained with ATCA about $1290$ days after optical discovery (see right panel of Fig.~\ref{fig: AMI_ATCA}). It shows an optically thin spectrum from $4.73$ GHz to $17.7$ GHz. The radio flare observed by \cite{cendes_2022} for AT 2018hyz is shown in Fig. \ref{fig: tde_comparison} together with the $15.5$ GHz light curve we first report here, and other known TDEs with delayed radio flares.

Next, we fit a power law function to our $15.5$ GHz light curve. We use \textsc{emcee} \citep{foreman_mackey_2013} to perform Markov chain Monte Carlo (MCMC) analysis to determine the posterior probability distributions of the parameters of the fitted model (and use flat priors). Based on the results of our fit  we conclude that the $15.5$ GHz flux density rises as $t^{\alpha}$ where $\alpha = 4.2 \pm 0.9$. Previous results showed a rise of $\sim t^{4.8}$ in the $3$ GHz band, and $\sim t^6$ in the $6$ GHz band \citep{cendes_2022}. Overall, we find that the steep rise in the $15.5$ GHz flux density matches the observed power laws in other bands even at later times (up to $\sim 1580$ days after optical discovery).

Based on their analysis of the radio emission, \cite{cendes_2022} conclude that the explanation of the late-time radio flare is a delayed launch of an outflow, about $\sim 750$ days after optical discovery. They also concluded that the steep rise in the radio flux density rules out any scenario of an outflow launched at the time of disruption (e.g., an off-axis jet, or a sudden increase in the ambient density). However, as discussed previously, a recent result by \cite{matsumoto_2023} suggests that an off-axis jet is possible and even likely. By generalizing the equipartition method to relativistic off-axis viewed emitters, \cite{matsumoto_2023} showed that the delayed emission from AT 2018hyz can result from an off-axis jet that was launched at the time of the TDE. Their analysis predicts that given the observed evolution of the peak flux density of $F_{\rm p} \sim t^5$, the radio light curve will peak around $3000$ days after optical discovery. 

\cite{Beniamini2020} developed an off-axis jet model to explain X-ray plateaus seen in GRB afterglows as afterglow emission from the jet core, as seen by an observer that is slightly misaligned with it. The model was extended later by \cite{duque_2022} and used to explain the rapid flaring activity in many GRB X-ray afterglows as the signature of the off-axis prompt emission on the same observer. Recently, \cite{Beniamini_2023_J1644} showed, using this model, that an off-axis jet with an opening angle of $\sim 21^{\circ}$, and a viewing angle of $\sim 9^{\circ}$ above the edge of the core, accounts for the radio emission observed from Swift J1644$+$54. Motivated by this, and the result of \cite{matsumoto_2023}, we make use of the entire radio observations published so far to model the delayed radio emission from AT\,2018hyz as arising from a narrow, powerful, off-axis jet.

\begin{figure*}
\includegraphics[width = \textwidth]{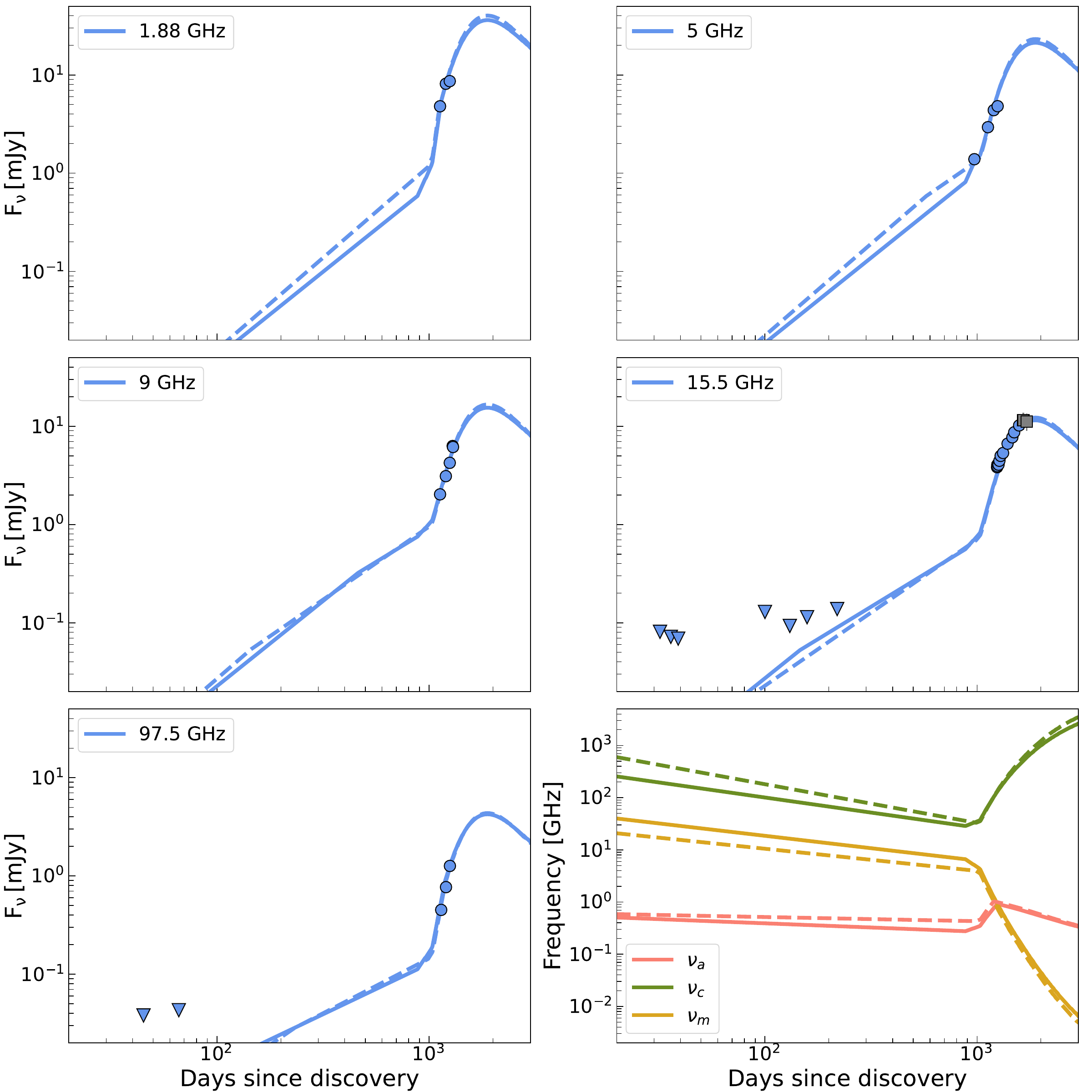}
    \caption{The solid lines are a relativistic off-axis forward shock model for AT 2018hyz described by $E_{\rm k,iso}=9.5 \times 10^{54}\mbox{ erg}, k=0.95, p=2.08, \epsilon_{\rm B}=0.10, \epsilon_{\rm e}=0.18, n(10^{19}\mbox{ cm})=0.016\mbox{ cm}^{-3}, \theta_0=0.12 \mbox{ rad}, \Delta \theta \equiv \theta_{\rm obs}-\theta_0=0.74 \mbox{ rad}, \Gamma_0>10$. Panels 1-5 depict the light curve at different observed frequencies. Measured values (and their errors) are shown in blue circles and $3\sigma$ upper limits are denoted by triangles. Panel 6 shows the evolution of the synchrotron characteristic frequencies  with time (see e.g. \citealt{granot_2002} for their definitions) for this model. The dashed lines are the same model but assuming $\epsilon_e = \epsilon_B$ which results in $E_{\rm k,iso}=1.3 \times 10^{55}\mbox{ erg}, k=0.85, p=2.12, \epsilon_{\rm B}=\epsilon_{\rm e}=0.1, n(10^{19}\mbox{ cm})=0.016\mbox{ cm}^{-3}, \theta_0=0.11 \mbox{ rad}, \Delta \theta \equiv \theta_{\rm obs}-\theta_0=0.74 \mbox{ rad},\Gamma_0>9.5$. The square grey data points were observed after finalizing the modelling. Plotted here is the fit without these points. These two points show the beginning of a predicted flattening of the light curve and indicate that we begin to observe the core of the jet.}
    \label{fig:18hyzmodel1}
\end{figure*}

\subsection{A narrow and powerful off-axis jet}
\label{subsec: jet_modeling}

Our modelling of AT 2018hyz follows the same procedure described in \cite{Beniamini_2023_J1644} for Swift J1644$+$57. We describe below some of the key features of this modelling and refer the reader to that work for the full details.

We consider, for simplicity, a `top-hat' jet, that consists of a core with uniform kinetic energy per unit solid angle, ${\rm d}E_{\rm k}/{\rm d}\Omega=E_{\rm k,iso}/4\pi$ that extends up to a finite latitude $\theta_0$. The jet is launched with a Lorentz factor, $\Gamma_0$. It propagates into an external medium that is characterized by a radial density profile $\rho =A r^{-k}$. The interaction between the jet and this surrounding material decelerates the jet and shocks the external medium. The dynamical evolution of the jet's Lorentz factor, $\Gamma(t)$, and radius, $R(t)$, can be split into four regimes: pre-deceleration, post-deceleration, post-jet break, and post-non-relativistic transition. The evolution within each segment is well approximated by a power-law (PL) expression, leading to a broken PL description of $R(t),\Gamma(t)$ at a general time, $t$. Once the dynamical evolution is known, one can express the number of emitting electrons, $N(t)$, the magnetic field, $B(t)$, and the minimum Lorentz factor, $\gamma_{\rm m}(t)$, above which electrons are accelerated to a PL of the form ${\rm d}N/{\rm d}\gamma\propto \gamma^{-p}$. This allows a full description of the synchrotron and synchrotron self-Compton (SSC) spectrum and peak flux evolution with time as seen by an observer viewing the jet `on-axis'. An extension to an off-axis observer can then be simply obtained with the use of only one extra parameter, $\theta_{\rm obs}$ (see e.g. \citealt{Beniamini_2023_J1644} for details).

Altogether, the model involves nine input physical parameters, allowing us to calculate a self-consistent flux evolution as a function of time and frequency. The model presented here demonstrates that an off-axis jet can explain the delayed radio flare observed in AT\,2018hyz. However, the physical parameters describing the system are not determined uniquely by the data. Therefore, we next present two probable sets of parameters to describe the radio emission under our framework. One for general, constant, micro-physical parameters, $\epsilon_{\rm e}$ and $\rm \epsilon_{\rm B}$, and the other while imposing the same assumption made in previous works $\epsilon_{\rm e} = \epsilon_{\rm B} = 0.1$ \citep{cendes_2022, matsumoto_2023}.

\begin{figure*}
\includegraphics[width = \textwidth]{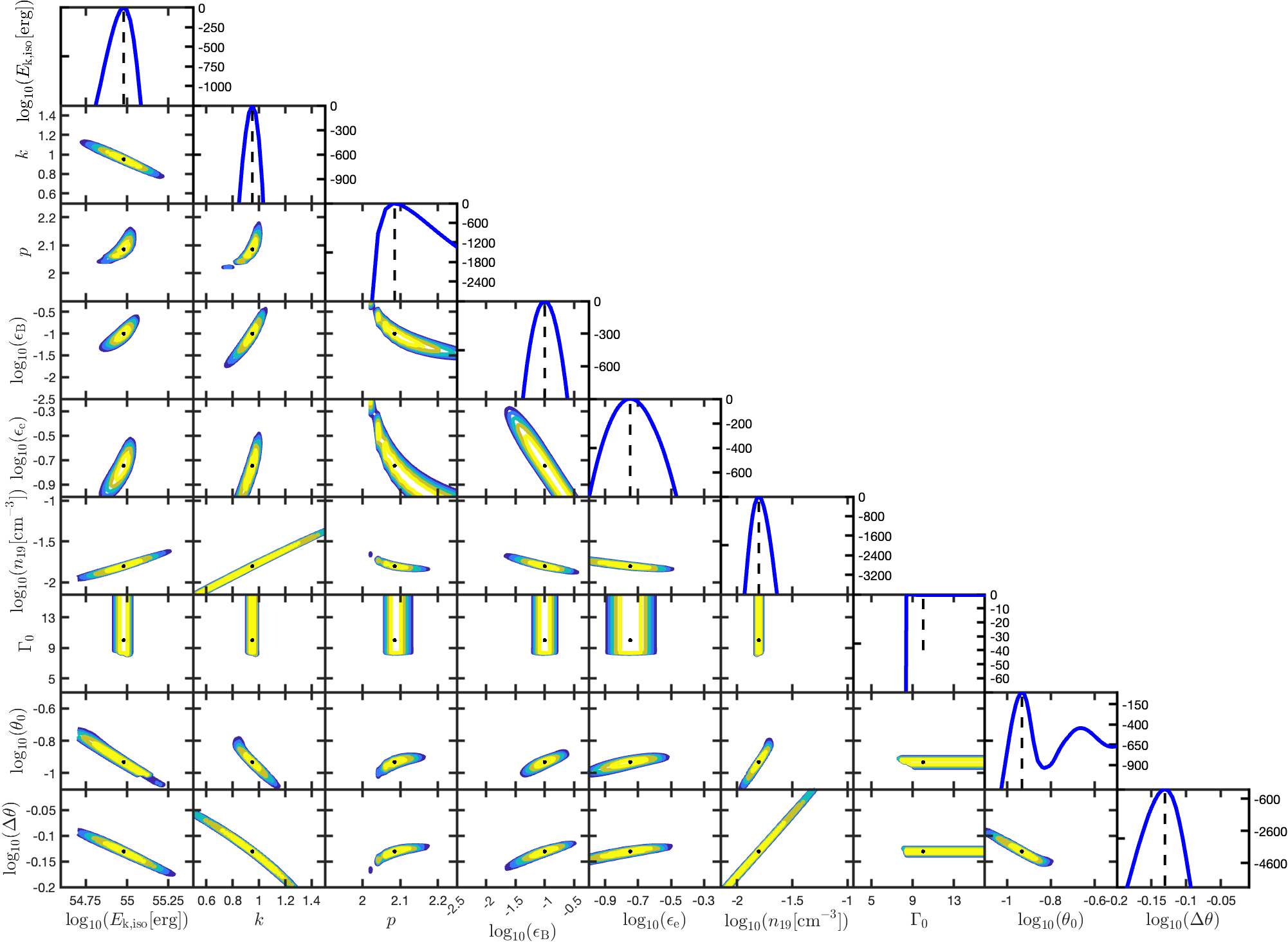}
    \caption{Correlations and marginalized log-likelihood from our multi-wavelength modelling of AT 2018hyz depicted in Fig. \ref{fig:18hyzmodel1}. Black dots and dashed lines denote our best-fit parameters. Contour lines represent marginalized likelihoods of: $0.9$,$10^{-2}$,$10^{-4}$,$10^{-6}$.}
    \label{fig: corner}
\end{figure*}

We perform an MCMC analysis to find the best-fit parameters for our model. The results of our analysis for AT 2018hyz are shown as solid lines in Fig. \ref{fig:18hyzmodel1}. The best-fit parameters are: $E_{\rm k,iso}= 9.5 \times 10^{54}\mbox{ erg},k=0.95, p=2.08, \epsilon_{\rm B}=0.10, \epsilon_{\rm e}=0.18, n(R_{\rm 0})=0.016 \mbox{ cm}^{-3}, \theta_{\rm 0}=0.12 \mbox{ rad},\upDelta \theta \equiv \theta_{\rm obs}-\theta_{\rm 0}=0.74 \mbox{ rad}$. Where we define a radius $R_{\rm 0} = 10^{19}$ cm in which we evaluate the number density $n(R_{\rm 0})$. We do not report here the $1 \sigma$ uncertainties obtained in the MCMC analysis (and presented in the MCMC corner plot; see Fig. \ref{fig: corner}) as they are the error on each parameter while keeping all the other parameters fixed. This analysis ignores the degeneracies between parameters in our model, and therefore, does not represent the true model uncertainties. We further discuss some of the degeneracies in our model later in this section. 

\begin{figure}
\includegraphics[width = \linewidth]{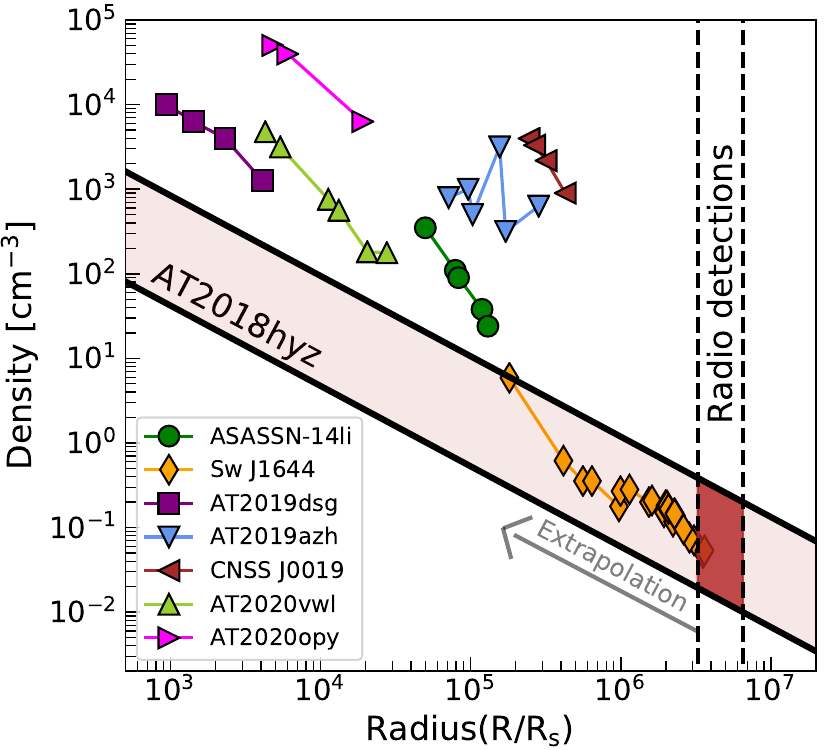}
    \caption{Density profiles around known TDEs (Sw J1644; \citealt{eftekhari_2018}, ASASSN-14li; \citealt{alexander_2016}, AT\,2019dsg; \citealt{stein_2021}, AT\,2019azh; \citealt{goodwin_2022}, CNSS J0019; \citealt{Anderson_2020}, AT\,2020opy; \citealt{Goodwin_2023a}, and AT\,2019vwl; \citealt{Goodwin_2023b}) based on their equipartition analysis, together with the range of density profile we infer from our off-axis jet modelling (shown red region). The dark-red region represents the radii we probe with the radio observations, and the shaded region is an extrapolation of the model to other regions.}
    \label{fig: density}
\end{figure}

The available data does not impose any upper limit on $\Gamma_0$ (recall that once the material decelerates the evolution of the spectral flux is independent of $\rm \Gamma_0$) and is consistent as long as $\rm \Gamma_0 > 10$. The model requires a large viewing angle to account for the sharp rise of the flux density at late ($\sim 3$\,yr) times. This, in turn, increases the energy requirements, since at times before the peak of an off-axis jet, a large fraction of the available energy is obscured from the observer due to relativistic beaming. We note that while the isotropic equivalent kinetic energy appears to be large, the collimated corrected energy, $E_{\rm \theta} \sim 0.5 \theta_0^2 E_{\rm k,iso}\approx 7\times 10^{52}\mbox{ erg}$ is within the energy budget for a TDE \citep{piran_2015,stone_2016}. Other parameters are consistent with expectations based on the modelling of other relativistic TDE jets and the modelling of GRB afterglows. Figure \ref{fig: density} shows a comparison of the fitted density structure to the density structure of the surroundings of other known TDEs.

\begin{figure}
\includegraphics[width = \linewidth]{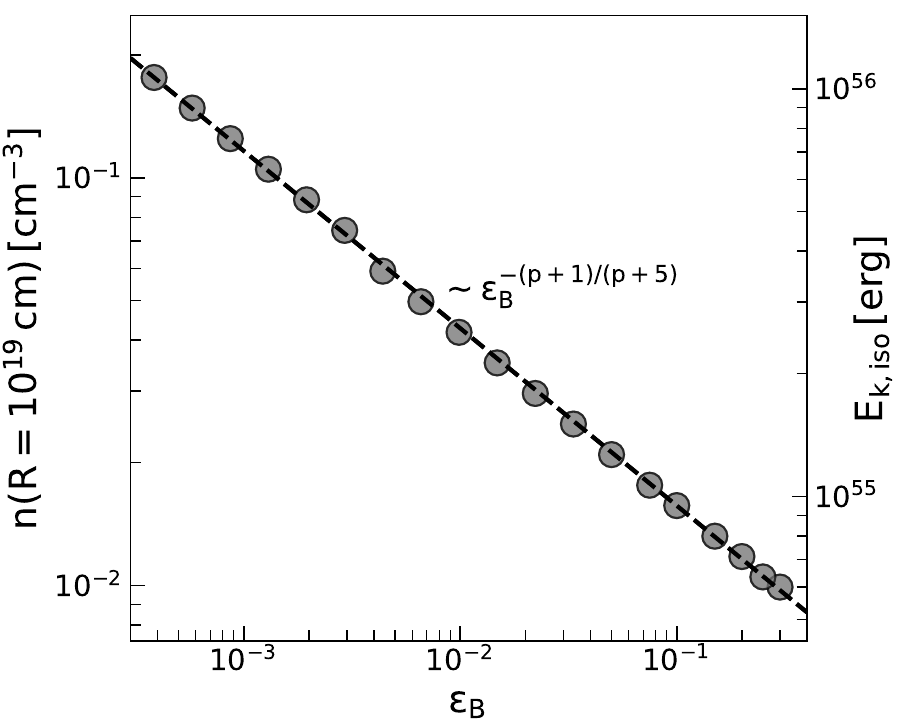}
    \caption{Different combinations of our model parameters that do not change the quality of our fit. Here we vary the density, $n \left(R_{\rm 0} \right)$ (proportional to $A$), and $\epsilon_{\rm B}$ while keeping all other parameters fixed, except for $E_{\rm k,iso}$ that we vary in proportion to the density. This illustrates three-parameter degeneracy discussed in \S\ref{subsec: jet_modeling}. The dashed line shows the analytical power-law for reference, where we used $p=2.08$ (based on our best-fitted model) which leads to $\sim \epsilon_{\rm B}^{-0.44}$.}
    \label{fig: degeneracy}
\end{figure}

As mentioned above the parameter space is degenerate. The underlying degeneracies involve combinations of three or more parameters. A notable example is the 3-parameter degeneracy between $A$, $E_{\rm k,iso}$ and $\epsilon_{\rm B}$. Specifically, changing $A\to XA$, $E_{\rm k,iso}\to X E_{\rm k,iso}$ and $\epsilon_{\rm B} \to X^{-(p+5)/(p+1)} \epsilon_{\rm B}$ leaves the quality of the fit approximately unchanged for a large range of values for $X$. This can be understood as follows. Applying the transformation above, $E_{\rm k,iso}/A$ remains constant. This in turn means that the radius and Lorentz factor at any time will also remain constant. In particular, this fixes all the dynamical time-scales of interest: deceleration, jet-break, non-relativistic transition and the time of the observed peak. Consider an observed frequency $\max(\nu_{\rm a},\nu_{\rm m})<\nu<\nu_{\rm c}$ (for our best fit model, most of the data is found to reside within this frequency range). The flux in this frequency range is given by $F_{\rm \nu}\propto F_{\rm max} \nu_{\rm m}^{(p-1)/2} \propto A^{(p+5)/4} \epsilon_{\rm B}^{(p+1)/4}$ (where we have ignored any dependence on parameters that are unchanged by the transformation). In particular, we see that after applying the transformation $F_{\rm \nu}$ remains unchanged. To show that this degeneracy is reflected in our numerical fits we present in Fig. \ref{fig: degeneracy} different combinations of the model parameters which produce a reasonable fit to the data. In practice, we vary the density, $n \left(R_{\rm 0} \right)$ (proportional to $A$), and $\epsilon_{\rm B}$ while keeping all other parameters fixed, except for $E_{\rm k,iso}$ that we vary in proportion to the density as explained above. 

Previous works \citep{cendes_2022, matsumoto_2023} considered equipartition ($\epsilon_{\rm e} = \epsilon_{\rm B}$). Since our model does not constrain $\epsilon_B$, we next repeat our analysis but now with fixing $\epsilon_{\rm e} = \epsilon_{\rm B} = 0.1$. The results are shown in dashed lines in Fig. \ref{fig:18hyzmodel1}. The best-fit parameters are: $E_{\rm k,iso}=1.3 \times 10^{55}\mbox{ erg}, k=0.85, p=2.12, n(R_{\rm 0})=0.016 \mbox{ cm}^{-3}, \theta_{\rm 0}=0.11, \upDelta \theta \equiv \theta_{\rm obs}-\theta_{\rm 0}=0.74$. The collimated corrected energy in that scenario is $E_{\rm k} \sim 8 \times 10^{52} \, \rm erg$. As before, the available data does not impose any upper limit on $\Gamma_{\rm 0}$ and it is consistent with the model as long as $\Gamma_{\rm 0} > 9.5$. As seen from Fig. \ref{fig:18hyzmodel1}, the difference between the best-fitted model described earlier and the model assuming equipartition is negligible and both models provide a good description of the radio emission seen from AT 2018hyz.

Overall, an off-axis jet describes well the radio emission from AT 2018hyz, however, the jet physical parameters are not uniquely determined by the set of radio observations available so far. We find that the minimal possible energy which provides a good fit is $E_{\rm k,iso} \geq 5 \times 10^{54} \, \rm erg$. This is equivalent to a limit on the collimated corrected energy of $E_{\rm k} \geq 3 \times 10^{52} \, \rm erg$, approximately similar to the jet energy interpreted for Swift J1644 by \cite{Beniamini_2023_J1644} under the same framework of an off-axis jet. \cite{matsumoto_2023} predict, for a viewing angle of $\pi/2$, that the emission will peak at radio wavelengths around $3000$ days after the disruption, and that the radio-emitting source moves $\sim 0.3 \, \rm mas \, yr^{-1}$. Based on the results of our analysis we predict that the radio emission will peak around $1900$ days. After finalizing the modelling we obtained two additional observations with AMI-LA. Our latest observations (marked as grey squares in Fig. \ref{fig:18hyzmodel1}) are in agreement with the best-fitted model that was based just on the earlier point, and show the possible onset of the light curve flattening, suggesting that we begin to see the core of the jet. The off-axis model also predicts the proper motion of the source. The movement of the source during the first $1000$ days is $\sim 3$ mas, and between $1000$ to $2000$ days, we expect a centroid movement of $0.1-0.5$ mas, depending on the exact form of spreading that is assumed. Thus, VLBI observations can confirm the scenario of an off-axis jet, or alternatively, rule out regions in the parameter phase space of such possible jets.

\section{Summary and conclusions}
\label{sec: summary}

Previously reported broadband, multi-timescale, radio observations of the TDE AT 2018hyz revealed a bright, late-time, and delayed, radio flare \citep{cendes_2022}. We present here follow-up observations of this TDE conducted with the AMI-LA and ATCA at radio wavelengths. The broadband observation obtained with ATCA about $1290$ days after optical discovery revealed an optically thin emission from $4.7$ to $17.5$ GHz. Our $15.5$ GHz light curve obtained with AMI-LA shows early limits up to $\sim 220$ days after optical discovery and a delayed flare that rises as $t^{4.2 \pm 0.9}$ up to $\sim 1580$ days after optical discovery.

We use all the radio data available to date to model the radio emission as arising from a jet that was initially observed off-axis. As the jet decelerates, the relativistic beaming cone becomes wider and it intersects the line of sight to the observer. We find an initially off-axis jet with a viewing angle of $\sim 42^{\circ}$ above the jet core, an opening angle of $\sim 7^{\circ}$, and an (isotropic equivalent) kinetic energy of $\sim 10^{55}$ erg~s$^{-1}$ that travels in a density profile of $\sim r^{-0.95}$. Extrapolating this model to later times predict that the radio light curves will peak at $\sim 1900$ days after optical discovery. We emphasize here that the framework of this model involves 9 free parameters and, naturally, the parameter phase space has degeneracies. $\epsilon_{\rm B}$, for example, is free to vary over 4 orders of magnitude, and the jet energy is only constrained within one order of magnitude (while satisfying $E_{\rm k,iso} \geq 5 \times 10^{54}$ erg). Furthermore, the density of the surrounding medium, assuming a single power-law structure, $r^{-k}$, is limited to $0.4 < k < 1.5$ while the density itself varies by a factor of $\sim 5$. Therefore, any discussion on the exact parameters and their uncertainties should be taken with care. However, it is clear that the delayed radio emission from AT 2018hyz can be explained by an off-axis jet that enters our line of sight at late times.

While the number of TDEs with observed late-time, delayed, radio flares is rising, it is still unclear if the underlying mechanism behind these flares is the same for all TDEs. The radio emission from AT 2018hyz is consistent with an off-axis jet. A similar off-axis signature has also been observed in Swift J1644$+$57 \citep{Beniamini_2023_J1644}. However, the delayed radio flare seen in ASASSN-15oi, for example, cannot be explained by such a model. In this latter case, it is possible that the delayed flare is a result of a delayed outflow ejection \citep{horesh_2021a}. Such a delayed outflow ejection, perhaps due to a transition in the accretion state of the SMBH, may also explain the radio–X-ray delayed flares in the TDE AT 2019azh (see \citealt{sfaradi_2022}). In order to reveal the nature of such late-time, delayed, radio flares we need late-time multi-frequency high-cadence radio observations as they provide insight into these fast-evolving flares (see \citealt{sfaradi_2022}).

\section*{Acknowledgements}
A.H. is grateful for the support by the I-Core Program of the Planning and Budgeting Committee and the Israel Science Foundation, the support by ISF grant 647/18, the United States-Israel Binational Science Foundation (BSF), and the support by ISF grant No. 2018154. We acknowledge the staff who operate and run the AMI-LA telescope at Lord's Bridge, Cambridge, for the AMI-LA radio data. AMI-LA is supported by the Universities of Cambridge and Oxford, and by the European Research Council under grant ERC-2012-StG-307215 LODESTONE. PB's research was supported by a grant (no. 2020747) from the United States-Israel Binational Science Foundation (BSF), Jerusalem, Israel. TP is supported by an Advanced ERC grant MultiJets. Parts of this research were conducted by the Australian Research Council Centre of Excellence for Gravitational Wave Discovery (OzGrav), project number CE170100004. The Australia Telescope Compact Array is part of the Australia Telescope National Facility which is funded by the Australian Government for operation as a National Facility managed by CSIRO. We acknowledge the Gomeroi people as the Traditional Owners of the Observatory site.

\section*{Data Availability}

The data underlying this article will be shared on reasonable request to the corresponding author.

\bibliographystyle{mnras}
\bibliography{main}
\bsp
\label{lastpage}
\end{document}